\documentclass[aps,prl,twocolumn]{revtex4-1}
\usepackage{graphicx}
\usepackage{natbib}
\usepackage{amsmath}
\newcommand{\be}{\begin{equation}}
\newcommand{\ee}{\end{equation}}
\begin{document}
\title{Student Variability in Learning Advanced Physics}
\author{T. Sampson, and M. Hilke}
\affiliation{Department of Physics, McGill University, Montr\'eal, Canada H3A 2T8}
\date{\today}

\begin{abstract}
Learning of advanced physics, requires a combination of empirical, conceptual and theoretical understanding. Students use a combination of these approaches to learn new material. Each student has different prior knowledge and will master new material at a different pace. However, conventional classroom teaching usually does not accommodate the different learning paces of students. To both, study and address this issue, we developed an iterative Online Learning Machine (iOLM), which provides new learning content to each student based on their individual learning pace and tracks their progress individually. The iOLM learning module was implemented using server side web software ({\em php}) to supplement the undergraduate course in electromagnetic waves for majors in physics in their second year. This approach follows the hybrid online learning model. Students had to complete a section of the course using iOLM, which was only presented online. The data obtained for this class showed a wide spread of learning paces, ranging from 0.1 to 0.5, where 1 is the maximum pace allowed by iOLM and 0 the lowest. The mean was $\mu$=0.25, with a standard deviation of $\sigma$=0.12. While the pretest showed a positive correlation between the student's pace and performance, the postest had zero correlation, indicating that giving more time and content to weaker students allows them to catch up.

\end{abstract}
\maketitle

\section{Introduction}

When a teacher is presented with a classroom full of students, a question which immediately arises  is: how does {\em every} student get to learn in a classroom setting, considering that each student has a different learning rate? Indeed, most learning theories suggest that the rate of learning depends on the student's prior knowledge. Going back to Aristotle, as described by Locke, the initial mind is like an empty disk, which is then imprinted with knowledge through experience and reasoning \cite{Locke1690}. New knowledge is further constructed from associations involving prior knowledge, which, in turns, affects the rate of learning \cite{Mayer96}. To deal with student variability, the model practiced by Socrates \cite{Hake92} involved one-on-one mentoring. More recently, Kuhn has suggested that scientific reasoning goes beyond inductive inference and follows a truth-seeking process that involves the coordination of theory and evidence in a social setting \cite{Kuhn04}. This process cannot be separated from prior knowledge. Recently, these different learning models have even been used to derive mathematical expressions describing student's rate of learning \cite{Pritchard08}.

Going back to Piaget, learning can be understood in terms of three reasoning levels, concrete operational, transitional, and formal operational \cite{Inhelder58}. Concrete operational reasoners use logic but cannot solve problems beyond a concrete context and have difficulties with abstract concepts and hypothetical tasks. Transitional reasoners fall between concrete and formal reasoning, where they find success with hypothetical tasks in some contexts. Formal operational reasoners can think abstractly and hypothetically and are able to synthesize available information to solve problems. This is typically associated with the scientific method with common use of hypothetical and deductive reasoning.

\section{Advanced Physics Learning}

To generalize these concepts to advanced physics students (beyond freshmen), where the students are already preselected based on their prior ability to solve scientific problems, we need to adapt Piaget's learning theory. We identified three main learning approaches: empirical, conceptual and theoretical. The empirical learning approach is closest in spirit to Piaget's concrete operational reasoning. A typical empirical approach will consist in discussing examples or demonstrations. In the context of interest here, i.e., reflection and refraction of an electromagnetic wave, this could be demonstrating the refracted angle in a particular experimental configuration and calculating the refracted angle using Snell's law. Here we just used the empirical learning approach to explain a concept.

The conceptual learning approach, is inspired by Piaget's transitional reasoning. In the context of refection and refraction, the conceptual approach might emphasize the notion of fermat's principle of least time, which can then be used to derive Snell's law and solve problems. The conceptual approach requires a more encompassing framework (like Fermat's principle) and the ability to use this framework to derive and solve problems.

The theoretical learning approach is closest to Piaget's formal operational reasoning, since it requires the highest degree of abstraction. In the example of reflection and refraction, the theoretical approach is best described by introducing Maxwell's equations and its use in the derivation of Snell's law. This approach requires the largest leap between the introduced content (a set of differential equations) and it's application to a simple problem (obtaining the angle of a reflected and refracted wave). Maxwell's equations constitute the overreaching theoretical framework, from which the rest can be derived and posses a higher level of abstraction then Fermat's principle, which can be described without equations.

To learn, most students use a combination of these 3 approaches in order to master a given topic. By mastering a topic, we mean the ability to solve a set of problems (success) as well as
the ability to explain the topic (understanding), which is harder to measure. If a student only learns using, for example, the theoretical approach, the student might not be able to solve a problem easier to solve with the conceptual approach. For instance, a student expert in solving Maxwell's equations might not understand some of its implications, which can be reflected in his or her inability to give an approximate solution to a problem where Maxwell's equations are difficult to solve, like in a situation where the interface is rough instead of flat. In this context, having learnt through the  theoretical approach does not necessarily imply an ability to solve problems, which might be more effectively solved at the conceptual or empirical level. Hence, while there is a hierarchy in terms of level of abstraction, there is no clear hierarchy in terms of outcome as measured by success.

\section{Reflection and Refraction}

In general, while there is abundant research on the teaching of electricity and magnetism (E\&M) at the introductory level \cite{introductory}, research on advanced E\&M is still fairly limited \cite{upper}. In this paper, we will analyze the learning of advanced E\&M, and in particular the topic related to reflection and refraction of an electromagnetic wave, or simply light, which can be mastered at different levels of expertise. In high school or freshmen physics the expertise is generally limited to the geometrical form of Snell's law, which interrelates the various angles of incidence ($\theta_i$), reflection ($\theta_r$) and refraction ($\theta_t$) as $n_1\sin\theta_i=n_2\sin\theta_t$ and $\theta_i=\theta_r$, where $n_1$ and $n_2$ are the indices of refractions of the incident and transmitted material, respectively. This form can be easily derived by applying Fermat's principle of least time. However, this approach largely limits the understanding to something which is purely geometrical in nature but nonetheless allows for a basic understanding of phenomena such as rainbows or distortions at the water/air interfaces. Let's call this expertise 1. In a more advanced calculus based description of refraction and reflection, students typically learn the origin of Snell's law in the context of Maxwell's equations and electromagnetic waves, which can be applied to interfaces and then provide general conditions for the electric and magnetic fields components at an interface. This approach requires a proper understanding of differential equations as well as surface and volume integrals. We can label this as expertise 2. A further approach, relevant to advanced undergraduate students, introduces the concept of complex angles as a way to account for the imaginary component of the index of refraction describing the absorbtion in a dielectric material (hereby referred to as expertise 3). Even more advanced topics on refraction, such as metamaterials are often left to the graduate level (or expertise 4).

Before taking the course, each student will have a distinct prior knowledge, which can be characterized by an initial expertise between 0 and 4 for this particular topic, where 0 corresponds to no prior knowledge (unlikely) and 4 to an advanced graduate knowledge. In an advanced university course setting, most students are expected to have an expertise of at least 1 when taking this course, based on course requirements. If every student had exactly the same prior knowledge one could expect a reasonably uniform learning pace following as based on existing learning theories \cite{Pritchard08}, which would validate the use of a single classroom for a uniform student cohort. As we will see in our data, this is not the case.

In our analysis, we are aiming for a final expertise 3 (about 1 year from graduation). In addition to the differences in expertise, we can divide the main topic (reflection and refraction) into several subtopics or levels of knowledge (1 to 5). Level 1 is the derivation and explanation of the electromagnetic fields at an interface; level 2 covers the reflection law and Snell's law; level 3 covers the angular dependence of the amplitude of a reflected and transmitted transverse magnetic wave; level 4 the transverse electric case and level 5 the reflection on a metal interface. Each topic is then divided further into the three main learning approaches (empirical, conceptual and theoretical).

We used the topic on reflection and refraction, since it nicely illustrates the various levels of understanding of the same topic, which immediately points us to the questions of how does a student go from an understanding of level A to level B? How different are the learning paths of Marie and Albert? How does the learning path from A to B influence the grasp of level B? Is it possible to adapt the learning path to Marie or Albert? In this work we provide a framework to address these questions. In pursuing this goal, we introduced an iterative learning tool, which we named iOLM for iterative Online Learning Machine and implemented it as part of a teaching module on reflection and refraction for a second or third year physics majors course at McGill on electromagnetic waves.

\section{Different learning paces}

So how different is Marie learning from Albert? While most studies seem to indicate that learning outcome is positively correlated with the time students spend engaged in learning \cite{Brown86}, it is debatable whether the variance of the learning outcome increases \cite{Brophy82} or decreases \cite{Bloom76,Bloom84} with more time spent learning. The situation is different though, if students spend different amounts of time engaged in learning. For instance, in reference \cite{Keith82} showed that homework has compensatory effects, in that a low ability student doing homework obtained equivalent grades to high ability students not doing homework. This seems to suggest that when weaker students spend more time engaged in learning, they can achieve a comparable outcome to stronger students. The inherent differences in learning abilities has led to a number of suggestions, including differentiated instruction, which was shown to lead to greater gains in mathematical understanding in K-12 students \cite{Subban06,Tomlinson03} as well as for students at the undergraduate level \cite{Chamberlin10}.

Following a similar reasoning, we designed a learning program, which adapts the learning pace and time to each individual student as measured by how many steps are required to go from level A to level B. In order to implement this program we chose to use a web format, since it offers the greatest flexibility to the students in terms of when and where they engage in the learning process. Simultaneously, it allows us to monitor the progress of each individual student, thanks to web-based database management. Confidentiality issues can be handled efficiently by using secure cryptography or anonymous IDs.

\section{Online instruction}

Online instruction has become increasingly widespread, particularly with university level courses. The main essence from previous studies is that student's learning outcome in online courses is equivalent to classroom instruction \cite{Jahng07,Means10,Zhao05,Sitzmann06}. However, students who register to online courses are less likely to complete their courses \cite{Beatty03,Moore03}.

While most online instruction models tend to focus on an all-online model, there are interesting results on hybrid models. Indeed, based on a large analysis commissioned by the U.S. Department of Education \cite{US09}, it was found that the positive outcome for online instruction was stronger for hybrid-online models than for all online models. The hybrid-online model was also found to be better than classroom instruction alone when additional material was included in the hybrid model but not in the classroom instruction. This strongly supports the idea of offering some  online content in the context of a traditional classroom course as is the case in the present study. In fact, our program iOLM, discussed in more detail below, was implemented as an add-on to a conventional classroom course and counted as an online homework. Interestingly, online homework was shown to lead to an increased outcome as compared to pencil homework \cite{Cheng04,Mestre02}.

\section{Description of iOLM}

The heart of iOLM was written using {\em php} code, which is a server side computing language mainly used for web interfaces and useful in dealing with databases. Using the {\em php} code, data was collected on the input of students who were presented with web-based quizzes in order to determine their mastery of the material corresponding to various levels of the topic on reflection and refraction of electromagnetic waves over the course of a two-week period. The main topic (reflection and refraction) was divided into 5 levels (subtopics), where each level had 3 additional sublevels (content and learning approaches).

\begin{figure}
\centering
\includegraphics[width=\columnwidth]{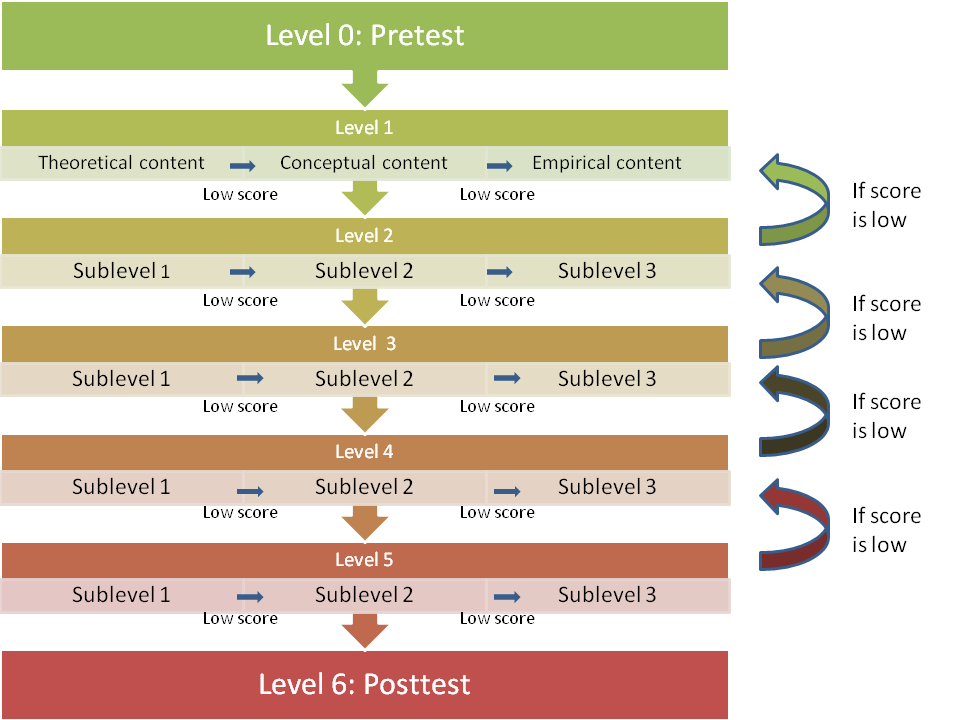}
\caption{The flow diagram of iOLM. Each arrow represents the outcome of a quiz (either success or no-success).}
\label{machine}
\end{figure}

The structure of iOLM is shown in figure \ref{machine} and was inspired by analogy to video games. This is interesting because it was shown that computer games in learning can be effective due to the additional emotional reactions in players \cite{Prensky01,Squire04}. The iOLM was structured as follows: First, an initial test (level 0) was administered in order to determine each student's individual prior knowledge with regards to the main topic. Next, the students were presented with a first content page related to level 1. Within each level, 3 sublevel content pages exist. Each sublevel content page uses a different learning approach (either empirical, conceptual or theoretical). After each content page a web-based quiz is administered based on a random selection of questions related to the corresponding level. The movement from one content page to the next was based on the success or not of each quiz. The algorithm used was the following: let $(n,m)$ describe level $n$ and sublevel $m$, then as long as $n>0$,

\be
\left\{
\begin{array}{rcll}
(n,m) & \rightarrow & (n+1,m) & \mbox{ if success}\\
(n,m) & \rightarrow & (n,m+1) & \mbox{ if no-success and }m\leq 2\\
(n,3) & \rightarrow & (n-1,1) & \mbox{ if no-success }
\end{array}
\right.
\ee
until $n=6$ is reached, which is the posttest page.

Success was defined as less than two incorrect answers for each quiz. For each level a student has three attempts to answer the quiz questions with at most two errors; upon each failure more supplementary material is provided on the webpage in addition to the material already presented. If nothing 'clicks' for the student within that time, then he or she must retake the previous level's quiz, and proceed once again. Upon completion of the last level's quiz, a final test (level 6) was administered which, upon comparison with the student's initial quiz score, can determine the improvement in the student's overall ability regarding this topic.

\section{Student learning trajectories}

Because the quizzes were web based, {\em php} was used to keep track of a variety of trends in the
students' performance. Notably, one can plot the current level to which a student has
progressed versus, say, the number of quizzes he or she has been administered. In this way
the 'trajectory' of the student can be plotted.

\begin{figure}
\centering
\includegraphics[width=\columnwidth]{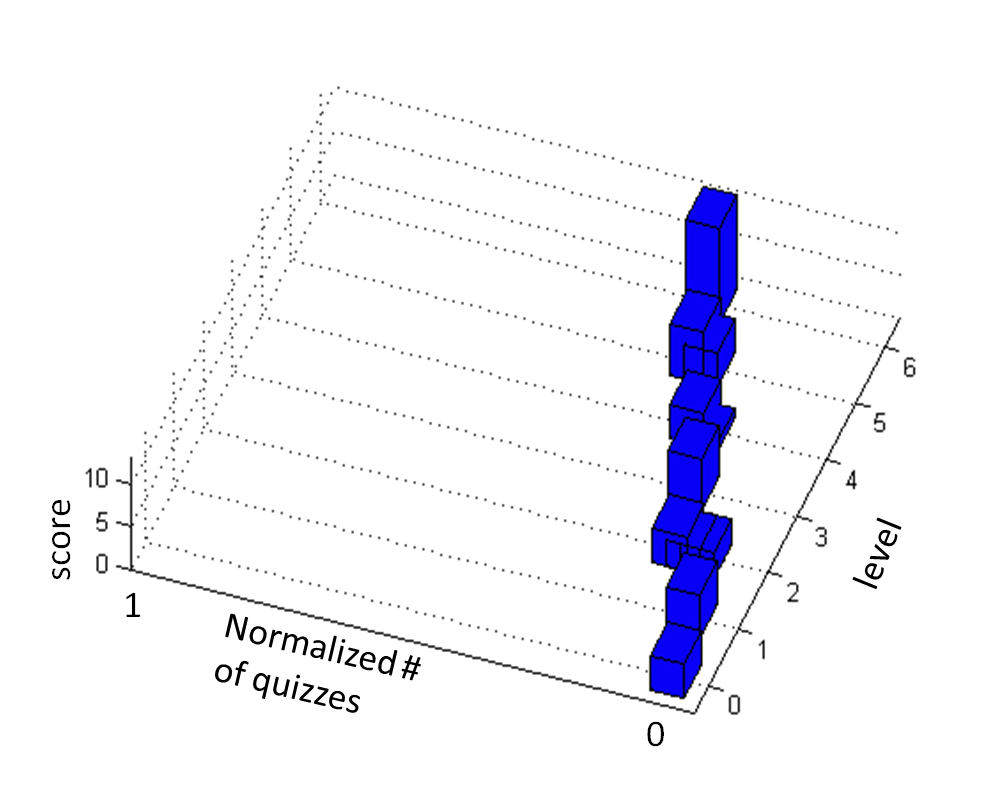}
\caption{A typical "fast" learning path. The score and the normalized number of trials is shown as a function of level.}
\label{fast}
\end{figure}

In figure \ref{fast}, a typical fast learning trajectory is shown. The score represents the number of correct answers to each quiz. The level 0 quiz corresponds to the initial pretest, used to measure prior knowledge on the topic. The level 6 quiz is the final quiz on the entire content from levels 1 to 5. The number of times a quiz was completed, was normalized to the maximum number of trials allowed (in this case 40). For each quiz, the questions were selected randomly from a large data set of questions corresponding to each level.

\begin{figure}
\centering
\includegraphics[width=\columnwidth]{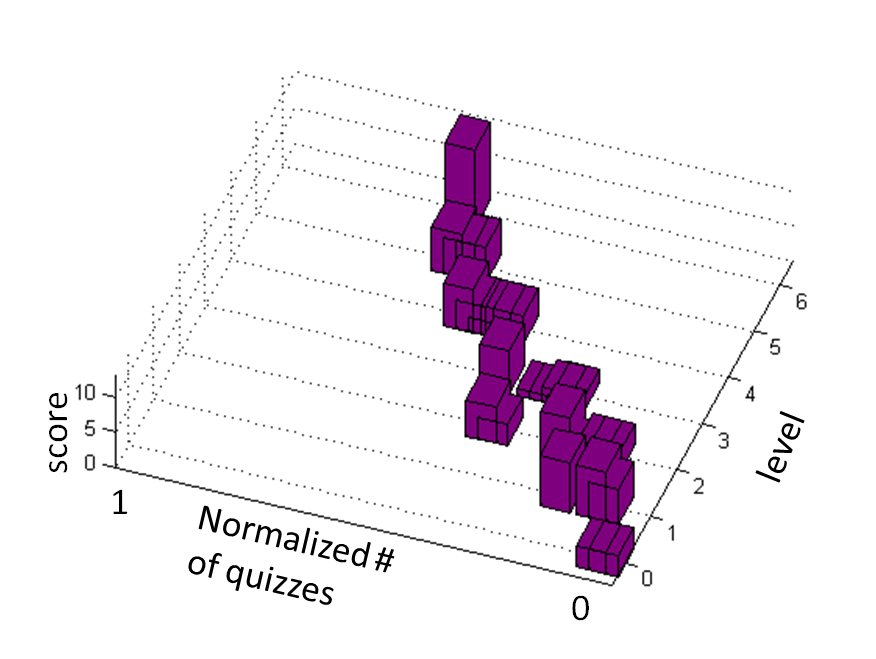}
\caption{A typical "average" learning path. The score and the normalized number of trials is shown as a function of level.}
\label{middle}
\end{figure}

In contrast, figure \ref{middle}, shows a typical average learning trajectory. Interestingly, one can see how the corresponding student went from level 2 back to level 1, before reaching level 3 and then back to level 2 before eventually reaching level 6.

\begin{figure}
\centering
\includegraphics[width=\columnwidth]{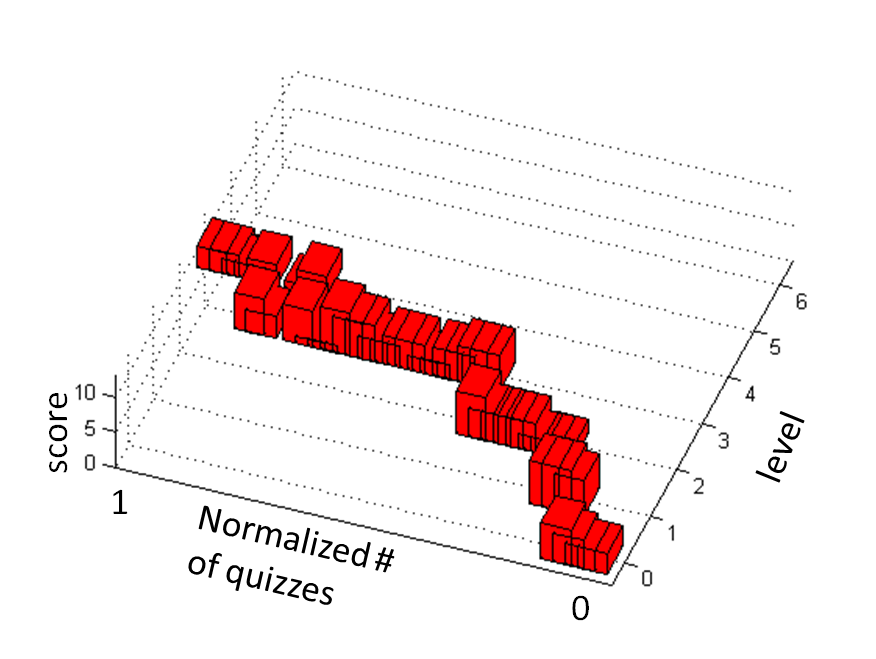}
\caption{A typical "stuck" learning path. The score and the normalized number of trials is shown as a function of level.}
\label{stuck}
\end{figure}

Some students also get stuck, like the one illustrated in figure \ref{stuck}. Here the student was not able to go beyond level 4 within the limit of 40 quizzes. For the students who got stuck, no final level 6 quiz data could be collected.

\begin{figure}
\centering
\includegraphics[width=\columnwidth]{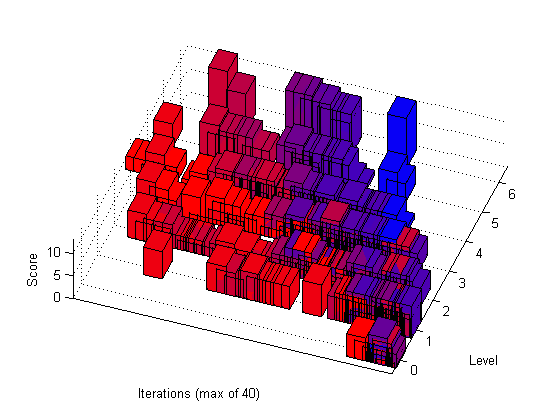}
\caption{All data from one class combined. The score and the normalized number of trials is shown as a function of level. The bluer the faster the learning pace. }
\label{all}
\end{figure}

Figure \ref{all} shows the collection of data representing all the data obtained in one class. The color coding was chosen as to reflect the typical pace of the student, with blue indicating a fast pace, while red indicating a slow pace.

\begin{figure}
\centering
\includegraphics[width=\columnwidth]{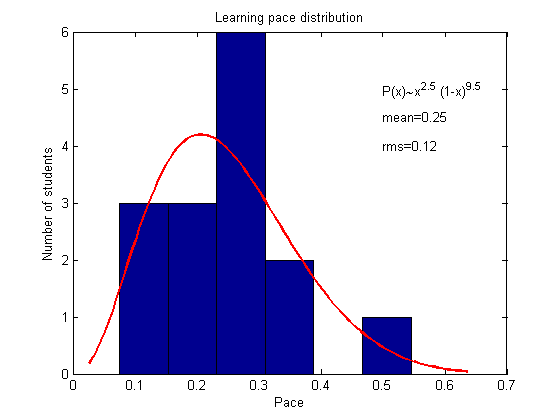}
\caption{Learning pace distribution. The number of students is shown as a function of learning pace, where 1 is the maximum learning pace allowed by iOLM and 0 the slowest. The distribution is fitted to a beta probability distribution, bound by 0 and 1. A mean of $\mu$=0.25 and a standard deviation of $\sigma=0.12$ was obtained.  }
\label{distribution}
\end{figure}

\section{Pace distribution}

from the different trajectories (see figure \ref{all}) it is now possible to extract the variability between students in terms of the learning pace. The distribution of the learning pace is shown in figure \ref{distribution}.

The differences in learning pace, range from 0.1 let's call him Albert to 0.5 (Marie) with a standard deviation of 0.12. The learning pace is defined as pace=6/(\# of quizzes). The average student needs 24 quizzes before reaching level 6. The most important result here is the large variability in learning pace between students. This immediately raises the question of how is it possible to teach both Albert and Marie effectively, considering that their learning pace is so different \cite{Tomlinson03}.

The advantage of a hybrid online module such as iOLM is to provide a framework, where each student can learn the material at his or her own pace. Hence Albert would see more content pages using different approaches, whereas Marie would see fewer content pages. For this implementation of iOLM, we chose to organize the 3 sublevels along a hierarchy based on level of abstraction. For sublevel 1 we used the theoretical approach or highest level of abstraction, for sublevel 2 the conceptual approach and for sublevel 3 the empirical approach. Marie would therefore be mainly  exposed to highly abstract material, whereas Albert would be provided with all three approaches. However, it is not always possible to draw a clear distinction between each sublevel based on one of these three learning approaches, since some levels are less prone to a clear distinction.

In some cases we used web applets for the conceptual approach. Applets or computer-assisted approaches have led to increased understanding of some physical concepts \cite{Hicks89,Perkins06}. The empirical approach was heavily based on worked-out examples. The division in many levels and sublevels also allowed for a more granular feed-back system to the students, which was shown to be the most important factor in computer assisted instructional approaches \cite{Vanlehn11}. After each quiz, the student is given the quiz score as well as the list of correct and wrong questions. However, when taking the quiz again for the same sublevel, both the questions and their order are changed by a random assignment of questions at the corresponding level.

\section{Outcome versus pace}

\begin{figure}
\centering
\includegraphics[width=\columnwidth]{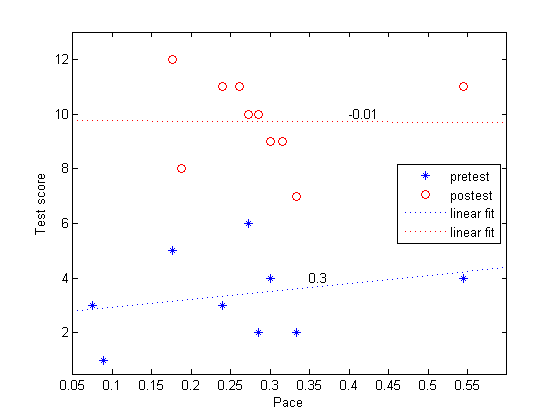}
\caption{The test score as a function of pace, for the pretest in blue and the posttest in red. the dotted line are the respective linear regressions.}
\label{correlation}
\end{figure}

The next question to arise quite naturally is the correlation between the learning pace and the learning outcome. In figure \ref{correlation}, we show the pretest and posttest test scores as a function of the student pace. Interestingly, while the pretest score correlates positively with the pace, the posttest score does not. This is consistent with the proposition that a student's learning pace correlates with the student's prior knowledge. For the slower students, after going through a learn process, which forces them to spend more time learning the same topic, this leads to an equalization of the learning outcome so that at the end students cannot be distinguished anymore based on their learning pace. Unfortunately, the number of students who participated in this study was too small to provide a statistical significant conclusion because the statistical error on the slope is too large. The goal for future work is to increase the sample size significantly by inviting more students at various universities to participate in iOLM. Nonetheless, the measured pace distribution is statistically significant and should be addressed by teaching techniques such as iOLM or differentiated instruction.

In general, a desirable feature of an iOLM - style program, or any pedagogical
method, for that matter, is adaptability; i.e. how well the course provides for slower
students whose needs may differ from those with a stronger grasp of the material. In terms
of statistics, this corresponds to a lack of correlations between the final test scores and the number of attempts students had to make in order to advance through the quiz levels. Arguably, this is the ideal outcome if the same mastery level of each students is desired. This can serve as a benchmark by which to measure the success of such an approach.

\section{Student reception and problems}

Overall, the reception by the students to introducing iOLM in their course was surprisingly neutral. New teaching methods often tend to be seen with a very critical eye. However, using an in-class clicker based survey we found that the overreaching response was neutral, i.e.,
half of the students found the experience interesting, and the other half not. 40\% would take another iOLM if given the choice, whereas 10\% would not and half were neutral. However, the learning outcome was evaluated as positive by only 15\% of the students. When the students were asked to put themselves in categories, 17\% put themselves in the fast category, 25\% in the slow and 58\% in the average category.

In terms of the feed-back specific to iOLM, some of the comments by the students in the course evaluations were very critical, with statements that they learned less than with "more standard approaches" or that they were not "fond of the innovated" teaching method, which illustrates the negative perception shared by some of the students on the structure. On the flip side, there were also many comments that were very supportive of "new ways of teaching" or the concepts that "couldn't have been learned from a book".

Some of the problems we encountered reduced the amount of data we could use. For instances, some
students have increased the apparent speed of their learning trajectory by creating 'practice runs', i.e. fake accounts used to practice the quizzes before proceeding through their 'final' trajectories having previously mastered the material. These 'final' trajectories had to be discarded, since they were meaningless. In addition, some pretest scores were marked as zeros, which may indicate that some students simply did not put effort into the preliminary exam; also some students accidentally submitted the same answers multiple times without meaning to. In these cases, the time stamp on the page would usually provide evidence of this and efforts were taken to ensure that the effect of these errors were removed.

\section{Conclusion}
Looking back at the questions we raised at the beginning, the first one is about how do student learn and how does prior knowledge affect the learning rate. Our results are consistent with most learning theories, which argue that prior knowledge influences positively the learning rate. This is indeed what we find as shown in figure \ref{correlation}, where pretest score correlate positively with learning rate. The ability to extract the different learning trajectories of all students was very instructive in this respect and illustrates the power of an iOLM-like learning environment, which provides detailed feed-back on each student.

Our model for learning of advanced physics was based on the assumption that there are three different main learning approaches (empirical, conceptual, and theoretical), which we implemented in iOLM. Unfortunately, we did not have enough data to analyze the effectiveness of each learning approach independently. This would require a much large sample size and a randomization of the order of each approach. This is an interesting extension planned for the next few years.

Finally, the question about what are the inherent differences between the learning rates of students in a typical higher level physics course, was effectively measured using iOLM. Indeed, the variability in the learning pace of each student and its distribution, where obtained and a mean of 0.25 was found with a standard deviation of 0.12, which means that approximately two thirds of the students fall within a pace of 0.13 and 0.37, while one third falls outside this window. This demonstrates the importance of finding instructional techniques, which can accommodate these different learning rates. An iOLM-like approach solves this by providing more learning material to slower students, thereby allowing them to catch-up as measured by the absence of a significant correlation at the completion of the iOLM exercise. Implementing such a tool online provides for an opportunity to be used by a vast community and thereby opens the door for gathering much more data with a very large sample size, which would enable the extraction of detailed correlations between different learning rates, approaches and outcomes.

\section{Acknowledgments}
We thank all the students in class {\em phys} 342 who agreed to participate in this study. We further acknowledge valuable feed-back from Jim Slotta and Yannis Dimitriadis.

\end{document}